\documentstyle[11pt]{article}

\def\be{\begin{eqnarray}}
\def\ee{\end{eqnarray}}

\def\Tr{{\rm Tr}\;}

\def\fm{{\rm fm}}
\def\thefootnote{\fnsymbol{footnote}}
\newcommand{\beq}{\begin{eqnarray}}
\newcommand{\eeq}{\end{eqnarray}}
\newcommand{\beqno}{\begin{eqnarray*}}
\newcommand{\eeqno}{\end{eqnarray*}}

\begin{document}
\begin{titlepage}\begin{center}
\hfill{ SNUTP-93-81}\\
\hfill{hep-ph/9401245}\\
\hfill{\it January 13, 1994}
\vskip 0.3in
{\LARGE\bf KAON-NUCLEON SCATTERING FROM CHIRAL LAGRANGIANS}
\vskip 0.4in
{\large Chang-Hwan Lee$^a$, Hong Jung$^a$, Dong-Pil Min$^a$
and Mannque Rho$^b$}\\
\vskip 0.1in
{\large a) \it Department of Physics and Center for Theoretical Physics,}\\
{\large \it Seoul National University, Seoul 151-742, Korea}\\
{\large b) \it Department of Physics, State University of New York}\\
{\large \it Stony Brook, N.Y. 11794, U.S.A.}\\ and\\
{\large \it Service de Physique Th\'{e}orique, CEA  Saclay}\\
{\large\it 91191 Gif-sur-Yvette Cedex, France\footnote{Permanent address.}}
\vskip 0.4in
{\bf ABSTRACT}\\ \vskip 0.1in
\begin{quotation}
The $s$-wave
$K^{\pm}N$ scattering amplitude is computed up to one-loop order corresponding
to next-to-next-to-leading order
(or N$^2$LO in short) with a heavy-baryon effective chiral Lagrangian.
Constraining the low-energy constants by on-shell scattering lengths,
we obtain contributions of each chiral order up to N$^2$LO and find that the
chiral corrections are ``natural" in the sense of viable
effective field theories.
We have also calculated off-shell $s$-wave $K^-N$ scattering amplitudes
relevant to kaonic atoms and $K^-$ condensation in ``nuclear star" matter
including the effect of $\Lambda (1405)$.
The $K^-p$ amplitude is found to be quite
sensitive to the intermediate $\Lambda (1405)$ contribution,
while the $K^-n$ amplitude varies smoothly with the C.M. energy.
The crossing-even
one-loop corrections are found to play an important role in determining the
higher-order chiral corrections.
\end{quotation}
\end{center}\end{titlepage}
\setcounter{footnote}{0}
\renewcommand\thefootnote{\#\arabic{footnote}}

Ever since Kaplan and Nelson\cite{KN} first predicted kaon condensation
in dense nuclear matter, there have been numerous investigations on this issue,
based both on
effective chiral Lagrangians\cite{BKR,PW,BKRT,lat,BLRT}
and on phenomenological off-shell meson-nucleon interactions
\cite{DEE,YNMK,lutz}.
The results have been quite confusing: While effective chiral Lagrangian
treatment leads generally to a robust behavior of the condensation phenomenon,
with a rather low critical density of order of three or four times ordinary
matter density $\rho_0$, the phenomenological treatment constrained by
empirical kaon-nucleon data seemed to predict otherwise, in some cases leading
to no possibility of condensation at all. This
raises the question as to whether the
chiral Lagrangian that predicts kaon condensation at low enough density is
consistent with empirical kaon nucleon and kaon nuclear interactions at low
energy. Inconsistency with scattering data would throw in doubt the premise
with which the predictions are made, thereby raising questions on the validity
of the arguments that go into the exciting scenario
of dense ``nuclear star" matter and of the formation of pygmy black holes,
recently put forward by Brown and Bethe \cite{brownbethe}.

In this paper, we address, using heavy baryon chiral perturbation
theory (HBChPT in short)\cite{Georgi,JM,Weinberg}, the question
as to whether the chiral Lagrangians used for
predicting kaon condensation are able to describe kaon nucleon and kaon nuclear
scattering. The advantage of HBChPT is that
it allows one to compute higher order chiral corrections systematically,
as shown recently in different contexts by several authors
\cite{JM,Weinberg,Meissner,PMR}. The recent paper by Brown, Lee, Rho and
Thorsson\cite{BLRT} (BLRT) addressed this problem to next-to-leading order in
chiral counting but this involved only the tree order. In this paper,
we will go one step further, that is to N$^2$LO,
by including one-loop contributions. Similar
one-loop calculations were recently reported by Bernard {\it et al.} on
s-wave pion-nucleon scattering\cite{bernard}. Focusing on
s-wave kaon-nucleon interactions as appropriate to the processes we are
interested in,
it will be shown that crossing-even one-loop
corrections will determine the size of higher-order
chiral corrections when the low-energy constants are constrained by
on-shell kaon-nucleon scattering  lengths. We also calculate
the off-shell $K^-N$ scattering amplitude and
find that our result is in good agreement with the recent phenomenological
fit of Steiner \cite{Steiner}.

Following Jenkins and Manohar\cite{JM}, we first write down the Lagrangian
that we shall employ. Let the characteristic momentum/energy scale that we are
interested in be $Q$. The standard chiral counting orders the amplitude as a
power series in $Q$, say, $Q^\nu$, with $\nu$ an integer. To leading order, the
kaon-nucleon amplitude $T^{KN}$ goes as ${\cal O}(Q^1)$, to next order as
${\cal
O}(Q^2)$
involving no loops and to next to next order ({\it i.e.}, N$^2$LO)
at which one-loop graphs enter as ${\cal O}(Q^3)$.
In terms of the velocity-dependent octet baryon fields $B_v$, the
octet meson fields exp$(i\pi_aT_a/f) \equiv\xi$,
the velocity-dependent decuplet baryon fields $T_v^{\mu}$,
the velocity four-vector $v_{\mu}$ and the
spin operator $S_v^{\mu}$ ($v\cdot S_v=0$,
$S_v^2=-3/4$),
the vector current $V_{\mu}=[\xi^{\dagger},\partial_{\mu}\xi]/2$ and the
axial-vector current $A_{\mu}=i\{\xi^{\dagger},\partial_{\mu}\xi\}/2$,
the Lagrangian density to order $Q^3$,
relevant for the
low-energy s-wave scattering, reads
\beq
{\cal L}^{(1)} &=&
 \Tr{\bar B }_v(iv\cdot{\cal D})B_v
     +2D\Tr{\bar B}_vS_v^{\mu}\{A_{\mu},B\}
      +2F\Tr{\bar B}_vS_v^{\mu}[A_{\mu},B]\nonumber\\
    &&-{\bar T}_v^{\mu}(iv\cdot{\cal D}-\delta_T)T_{v,\mu}
      +{\cal C}({\bar T}_v^{\mu}A_{\mu}B_v+\bar{B}_vA_{\mu}T_v^{\mu})
      + 2{\cal H}\bar{T}_v^{\mu}(S_v\cdot A)T_{v,\mu} \\
{\cal L}^{(2)} &=&
   a_1\Tr{\bar B}_v\chi_+B_v
   +a_2\Tr{\bar B}_vB_v\chi_+
   +a_3\Tr{\bar B}_vB_v\Tr\chi_+\nonumber\\
    &&+d_1\Tr{\bar B}_vA^2B_v+d_2\Tr{\bar B}_v(v\cdot A)^2B_v
    +d_3\Tr{\bar B}_vB_vA^2+d_4\Tr{\bar B}_vB_v(v\cdot A)^2\nonumber\\
  &&+d_5\Tr{\bar B}_vB_v \Tr A^2
  +d_6\Tr{\bar B}_vB_v\Tr (v\cdot A)^2
    +d_7\Tr {\bar B}_vA_{\mu}\:\Tr B_vA^{\mu}\nonumber\\
  &&+d_8\Tr {\bar B}_v(v\cdot A)\:\Tr B_v (v\cdot A),\\
{\cal L}^{(3)} &=&
c_1\Tr {\bar B}_v(iv\cdot{\cal D})^3B_v
    +g_1\Tr {\bar B}_vA_{\mu}
      (iv\cdot\stackrel{\leftrightarrow}{\cal D})A^{\mu}B_v
 +g_2\Tr B_vA_{\mu}(iv\cdot\stackrel{\leftrightarrow}{\cal D})A^{\mu}{\bar B}_v
     \nonumber\\
 &&+g_3\Tr {\bar B}_vv\cdot A(iv\cdot\stackrel{\leftrightarrow}{\cal D})
     v\cdot AB_v
    +g_4\Tr B_vv\cdot A(iv\cdot\stackrel{\leftrightarrow}{\cal D})v\cdot A
     {\bar B}_v\nonumber\\
   && +g_5\left(\Tr  \bar B_v A_{\mu} \Tr (iv\cdot\stackrel{\rightarrow}
      {\cal D}) A^{\mu}B_v
      -\Tr \bar B_v A_{\mu}(iv\cdot \stackrel{\leftarrow}{\cal D})
     \Tr A^{\mu}B_v \right)\nonumber\\
   && +g_6\left(\Tr \bar B_v v\cdot A \Tr B_v (iv\cdot\stackrel{\rightarrow}
    {\cal  D})v\cdot  A
     -\Tr \bar B_vv\cdot A (iv\cdot\stackrel{\leftarrow} {\cal D}
      )v\cdot A \Tr B_v v\cdot A\right)\nonumber\\
  && +g_7 \Tr \bar B_v [v\cdot A,[iD^\mu,A_\mu]]B_v
 +g_8 \Tr B_v[v\cdot A,[iD^\mu,A_\mu]]\bar B_v\nonumber\\
  &&+h_1\Tr {\bar B}_v\chi_+
	  (iv\cdot{\cal D})B_v
      +h_2\Tr {\bar B }_v(iv\cdot{\cal D})B_v
	 \chi_+
  +h_3\Tr {\bar B }_v(iv\cdot{\cal D})B_v \Tr \chi_+\nonumber\\
 && +l_1 \Tr \bar B_v [\chi_-, v\cdot A] B_v
 +l_2 \Tr \bar B_v B_v [\chi_-, v\cdot A]
 +l_3[\Tr \bar B_v \chi_-,\Tr B_vv\cdot A],
 \eeq
where the covariant derivative ${\cal D}_{\mu}$ for baryon fields is defined
by
\beq
 {\cal D}_{\mu}B_v &=& \partial_{\mu}B_v+[V_{\mu},B_v], \nonumber\\
  {\cal D}_{\mu}T_{v,abc}^{\nu} &=& \partial_{\mu}T_{v,abc}^{\nu}+
	(V_{\mu})_a^dT^{\nu}_{v,dbc}+(V_{\mu})_b^dT_{v,adc}^{\nu}
     +(V_{\mu})_c^dT_{v,abd}^{\nu},
\eeq
$\delta_T$ is the $SU(3)$ invariant decuplet-octet mass difference,
and
\beq
 \chi_{\pm} &\equiv& \xi {\cal M} \xi {\pm}
 \xi^\dagger {\cal M} \xi^\dagger,
\eeq
with  ${\cal M}={\rm diag}(m_u,m_d,m_s)$
the quark mass matrix that breaks chiral symmetry explicitly. The constants
$D$, $F$, $a_i$,..., $l_i$ are determined as described below.
The decuplet fields are not written down in ${\cal L}^{(2),(3)}$ since they do
not figure to N$^2$LO in the s-wave amplitudes we are interested in.

Despite the awesome appearance of the Lagrangian density with its numerous
terms contributing, there is a large simplification for the s-wave $K^{\pm}N$
amplitudes. The subleading terms ({\it i.e.}, terms with $\nu\geq 2$) involving
the spin operator $S_{\mu}$ do not contribute to the $s$-wave
$K^{\pm}N$ amplitudes, since they are proportional to
$S\cdot q$, $S\cdot q'$, or $S\cdot q
S\cdot q'$, all of which vanish. What this means is that there is
no contribution to the $s$-wave meson-nucleon scattering
amplitude from one-loop diagrams in which the external meson lines couple to
baryon lines through the axial vector currents. Thus we are left with only
six topologically distinct one-loop diagrams, Fig.1, (out of thirteen in all)
for the $s$-wave meson-nucleon scattering apart from the usual
radiative corrections in external lines.
Since we are working to ${\cal O}(Q^3)$, only
${\cal L}^{(1)}$ enters into the loop calculation. Now ${\cal L}^{(2)}$
contributes terms at order $\nu=2$, receiving no contributions from the
loops, as discussed already in \cite{BLRT}.
These will be completely given by the $KN$ sigma term and calculable
$1/m_B$ corrections. The terms in ${\cal
L}^{(3)}$, numerous as they are, remove the divergences in the one-loop
contributions and supply finite counter terms that are to be determined
empirically. As we will mention later, these constants are determined solely by
isospin-odd amplitudes, the loop contribution to isospin-even amplitudes being
free of divergences.

As discussed in \cite{BLRT}, to order $Q^2$, only the tree diagrams
contribute. Here as a first analysis, we extend the argument of \cite{BLRT} by
incorporating $\Lambda (1405)$ which plays an important role in the $K^- p$
channel \cite{lambda1405}.
Let $q$ ($q'$) denote the four-momenta of the
incoming (outgoing) mesons and $v^{\mu}$ ($v^2=1$, $v^0>0$) be
the  velocity four-vector of the nucleons in the limit $m_N\to\infty$.
The velocity four-vector of the nucleon assumed to be ``heavy" does not
change throughout the
meson-nucleon scattering as long as the momentum transfer is small
enough compared with the nucleon mass \cite{Georgi,JM}.
The on-shell
$s$-wave $K^{(\pm)}$ scattering amplitudes at tree level
can be readily written down from ${\cal L}^{(1)}$ and ${\cal L}^{(2)}$
with ${\hat m}=(m_u+m_d)/2$:
\beq
f^2\:T_v^{K^{\pm} p}&=&
   \mp{1\over 2}v\cdot(q+q')
-(a_1+a_2+2a_3)({\hat m}+m_s)\nonumber\\
&&  +{1\over 2}(d_1+d_3+2 d_5+d_7)q\cdot q'
   +{1\over 2}(d_2+d_4+2 d_6+d_8)v\cdot q v\cdot q',
 \label{TKp}\\
f^2\:T_v^{K^{\pm} n}&=&
   \mp{1\over 4}v\cdot(q+q')
-(a_2+2a_3)({\hat m}+m_s)\nonumber\\
 &&+ {1\over 2}(d_3+2 d_5)q\cdot q'
  +{1\over 2}(d_4+ 2 d_6)v\cdot q v\cdot q'.
  \label{TKn}
\eeq
Note that
the leading order contribution -- the first term in each amplitude --
comes from the nucleon coupling to the vector current and is odd under crossing
($q\leftrightarrow -q'$). As it stands,
it is attractive for $K^-p$ at threshold. If this were the main story, it would
be in
disagreement with the repulsion observed in nature\cite{D}. The cause for this
is well-known \cite{lambda1405}: The presence of $\Lambda (1405)$
slightly below the $K^-p$ threshold results in the repulsion for the
$s$-wave $K^-p$ scattering overcoming the vector attraction. The effect
of $\Lambda (1405)$ must therefore be taken into account.

Now how do we incorporate $\Lambda (1405)$ in chiral perturbation theory?
As discussed in \cite{BLRT}, since it is a bound state of the baryon-kaon
complex\cite{lambda1405}\footnote{It is a bound state of an $SU(2)$ soliton and
an s-wave kaon in the Callan-Klebanov model \cite{SMNR}.}, it should be
introduced as an independent heavy baryon field much like the resonance
$\Delta$ (more generally the decuplet $T_\nu^\mu$).
In particular it cannot be generated as a loop correction since the latter
would
involve ``reducible graphs" involving infrared singularities, rendering
the bound state inaccessible to chiral perturbation theory. Denoting
the $\Lambda (1405)$
by $\Lambda_R$, we can write the leading-order Lagrangian density as
\beq
{\cal L}_{\Lambda_R}&=&
   \bar\Lambda_R(iv\cdot\partial-m_{\Lambda_R}+m_B)\Lambda_R
   + \left( \sqrt{2} {\bar g}_{\Lambda_R}\:\Tr (\bar\Lambda_R v\cdot A B_v)
    + {\rm h.\:c.}\right) .
\eeq
At tree order, the $\Lambda (1405)$ contributes only to the $s$-wave $K^{\pm}p$
scattering:
\beq
 f^2\:T_{v,\Lambda_R}^{K^{\pm}p} &=&
  -{{\bar g}^2_{\Lambda_R}v\cdot q'v\cdot q\over
	       {1\over 2}[v\cdot(q-q')\mp v\cdot(q+q')]
	       +m_B-m_{\Lambda_R}}.
 \label{TLambda}
\eeq
Putting eqs.(\ref{TKp}),(\ref{TKn}) and (\ref{TLambda}) together,
we obtain the complete ${\cal O} (Q^2)$ ({\it i.e.}, tree-order)
s-wave $K^{\pm} p$ scattering lengths (see\cite{Hohler})
\beq
 a_0^{K^{\pm}p}&=& {m_N\over 4\pi f^2(m_N+M_K)}
		   \big[\mp M_K-{{\bar g}^2_{\Lambda_R}M_K^2\over
			 m_B\mp M_K-m_{\Lambda_R}}
			 +(\bar d_s+\bar d_v)M_K^2\big]
\nonumber\\
 a_0^{K^{\pm}n}&=& {m_N\over 4\pi f^2(m_N+M_K)}
		   \big[\mp{M_K\over 2}+(\bar d_s-\bar
d_v)M_K^2\big],\label{12}
\eeq
where we have decomposed the contribution of order $Q^2$ -- which is
crossing-even -- into a
$t$-channel isoscalar ($\bar d_s$) piece and an isovector ($\bar d_v$) piece:
\beq
\bar d_s \!\!\! &=&\!\!\! -{1\over 2B_0}(a_1+2a_2+4a_3)+
   {1\over 4}(d_1+d_2+d_7+d_8)+{1\over 2}(d_3+d_4)+d_5+d_6,\nonumber\\
\bar d_v \!\!\!&=&\!\!\! -{1\over 2B_0}a_1+{1\over 4}(d_1+d_2+d_7+d_8),
\label{dformulas}\eeq
with $B_0=M_K^2/(\hat m+m_s)$.

The empirical scattering lengths\cite{D,BS} are
\beq
  a_0^{K^+p}= -0.31\:\fm, && a_0^{K^-p}= -0.67+i\:0.63\:\fm \nonumber\\
  a_0^{K^+n}= -0.20\:\fm, && a_0^{K^-n}= +0.37+i\:0.57\:\fm.
\label{exp}
\eeq
The data for $K^{\pm}n$ are not well determined, so we cannot pin down the
parameters in a quantitative way. Nonetheless, we can use
these data to constrain the low-energy constants in our effective chiral
Lagrangian. To tree order, the amplitudes are real, the
imaginary parts of the amplitudes appearing at ${\cal O}(Q^3)$ involving loop
graphs. In eq.(\ref{12}),
$f$ is the meson decay constant in chiral limit and the difference between
$f_{\pi}$ and $f_K$ is of order $Q^4$. Therefore we are allowed to simply take
$f\approx f_{\pi}\approx 93$ MeV, the physical value.
Now requiring consistency with the data at tree
level leads to
\beq
\left(\bar d_s-\bar d_v\right)_{emp}&\approx& (0.05\sim 0.06)\: \fm,
\nonumber\\
\left(\bar d_s+\bar d_v\right)_{emp}&\approx&  0.13 \: \fm, \hspace{0.5in}
{\bar g}^2_{\Lambda_R}=0.15
  \label{Lambdaparameter}
\eeq
with $m_{\Lambda_R}=1.405\:{\rm GeV}$\footnote{
${\bar g}^2_{\Lambda_R}=0.15$ corresponds to $g^2_{\Lambda}/4\pi\approx 0.3$
in the conventional notation\cite{D}.}.
Let us call these ``tree-order empirical."
Although it is difficult to make a precise statement due to the uncertainty in
the data and the $\Lambda (1405)$ parameters, the above values provide a
persuasive indication that {\it
the net  contribution of order $Q^2$ or higher for $K^{\pm}p$ ($K^{\pm}n$)
is attractive and  amounts to $ \approx 33\:\%$ ($26\sim 31\:\%$) of the
strength given by the leading-order vector coupling.} This feature will be
reconfirmed later at one-loop order.

Before going to the next order that involves loops, we discuss briefly what one
can learn from the ``tree-order empirical" eq.(\ref{Lambdaparameter}).
The structure of the terms involved here is simple enough to render a
relatively unambiguous interpretation. For this, first we note that the
constants $\bar{d}_{s,v}$ consist of the $KN$ sigma term $\Sigma_{KN}
=-\frac 12 (\hat{m}+m_s)(a_1+2a_2+4a_3)$ involving the quark mass matrix
and the $d_i$ terms containing two time derivatives. As argued in \cite{BLRT},
the latter should be given by the leading $1/m_B$ correction in the
heavy-fermion formalism with no renormalization by loop graphs.
This can be readily seen from the chiral counting rule. The leading $1/m_B$
correction can be computed as explained in \cite{BLRT}
from kaon-nucleon Born diagrams of relativistic
chiral Lagrangians with the octet and decuplet intermediate states
by taking the limit $m_B\rightarrow \infty$. Assuming flavor
$SU(3)$ symmetry, one can calculate, in conjunction with the s-wave $\pi N$
scattering lengths \footnote{
The decuplet contributions were omitted in \cite{BLRT} without justification.
Here we rectify that omission and find that their contributions
are essential. In calculating the decuplet contributions to the lowest order in
$1/m_B$ from Feynman graphs in relativistic formulation, one
encounters the usual off-shell non-uniqueness characterized by a factor $Z$
\cite{benmerrouche} in the decuplet-nucleon-meson vertex
when the decuplet is off-shell. Our result corresponds to taking $Z=-1/2$
consistent with s-wave $\pi N$ scattering lengths at the same chiral order,
${\cal O}(Q^2)$.
}
\beq
\bar d_{s,\frac 1m}
 &=& -\frac {1}{48}\left[(D+3F)^2+9(D-F)^2\right]\frac{1}{m_B}
   -\frac {1}{6} |{\cal C}|^2 \frac{1}{m_B}
 \nonumber\\
  &\approx & -0.55 /m_B \approx - 0.115 \; \fm
 \nonumber\\
\bar d_{v,\frac 1m}
 &=& -\frac {1}{48}\left[(D+3F)^2-3 (D-F)^2\right]\frac{1}{m_B}
  +\frac {1}{18} |{\cal C}|^2  \frac{1}{m_B}
 \label{1/mtheory}\\
  &\approx & 0.057 /m_B \approx 0.012 \; \fm ,
\nonumber
\eeq
where the coupling $|C|^2$($\approx 2.58$) that appears in the decuplet
contributions was fixed\footnote{The decuplet-octet mass difference
$\delta_T$ figures only in the denominator $1/(m_B+\delta_T)$, so it
contributes at ${\cal O}(1/m_B^2)$ which we ignore here.}
from $\Delta(1230)\rightarrow N\pi$ decay width. We have used $D=0.81$ and
$F=0.44$ as determined at tree order from hyperon decays. The results
eq.(\ref{1/mtheory}) together with the empirical values
(\ref{Lambdaparameter}) imply that $\Sigma_{KN}\approx 2 m_\pi$ which is
consistent with a negligible strangeness content of the proton, $\langle
P|\bar{s}s|P\rangle\approx 0$.

Since the next-to-leading order (${\cal O} (Q^2)$) chiral corrections
are not small,
it is clearly important to go
to the next order in the chiral expansion.
At order $Q^3$, we have contributions from one-loop graphs given by Figure 1
and counter-term contributions
from ${\cal L}^{(3)}$. The ${\cal O} (Q^3)$ corrections involving $\Lambda
(1405)$ will be treated below. All the counter-term contributions to $s$-wave
$K^{\pm}N$ scattering are crossing-odd and have incalculable low-energy
constants.
The crossing-even terms coming from the one-loop graphs involving
${\cal L}^{(1)}$ are finite whereas the
crossing-odd terms from the same graphs are renormalized by the counter terms.
Thus the $s$-wave $K^{\pm}N$ scattering lengths coming from the order $Q^3$
terms can be written as
\beq
 \delta_3a_0^{K^{\pm}p} &=&
    {m_N\over 4\pi f^2 (m_N+M_K)}\big[(L_s+L_v)\pm(\bar g_s+\bar g_v)
    \big] M_K^3,\nonumber\\
 \delta_3a_0^{K^{\pm}n} &=&
    {m_N\over 4\pi f^2 (m_N+M_K)}\big[(L_s-L_v)\pm(\bar g_s-\bar g_v)
    \big]M_K^3.\label{delta32}
 \eeq
Here $L_s$ ($L_v$) is the finite crossing-even t-channel isoscalar (isovector)
one-loop contribution
\beq
  L_s M_K^3 &=& \frac{1}{128\pi f^2}  \Big\{
          +\frac{1}{3} (D-3F)^2 (M_\pi^2+3 M_\eta^2) M_\eta
 -9 M_K^2 \sqrt{ M_\eta^2-M_K^2}  \Big\}
\nonumber\\
 L_v M_K^3 &=& \frac{1}{128\pi f^2}\Big\{ -\frac 13 (D+F) (D-3F)
     (M_\pi^2+3M_\eta^2) (M_\pi+M_\eta),
\\
 && \;\;\:\;\;\;\;\;\;\; -\frac{1}{6} (D+F) (D-3F) (M_\pi^2+3M_\eta^2)
  (M_\pi^2+M_\eta^2)\int^1_0 \frac{1}{\sqrt{(1-x)M_\pi^2 +x M_\eta^2}}
 \nonumber\\
&& \;\;\:\;\;\;\;\;\;\;  -3 M_K^2\sqrt{M_\eta^2-M_K^2} \Big\}.\nonumber
\eeq
One can show that the baryon decuplets do not contribute to the s-wave
crossing-even amplitudes.\footnote{The diagrams (d), (e) and (f)
in Figure 1 do contribute to crossing-even amplitudes. However the
contributions of (d) and (f)
cancel each other between the decuplet- and octet-$K$, $\pi^\pm$ intermediate
states as shown in \cite{Montano}. The
non-vanishing contributions of (d) and (f) come
only from the octet-$\pi^0$,$\eta$ intermediate states. As for the
graph (e), only the octet baryons contribute.}
If we use
$D=0.81$ and $F=0.44$,
we obtain
\beq
  L_sM_K &\approx& -0.109\:\fm,\nonumber\\
  L_vM_K &\approx& +0.021\:\fm.
\eeq
The quantity $\bar g_s$ ($\bar g_v$) in eq.(\ref{delta32})
 is the crossing-odd t-channel isoscalar
(isovector) contribution from one-loop plus counter terms which after the
standard dimensional regularization, takes the form
\be
\bar{g}=\sum_i \gamma_i z^r_i (\mu)+\sum_i \left(\alpha_i {\rm ln}
\left(\frac{m_i}{\mu}\right) +f_i (m_i)\right)\label{scaleindep}
\ee
where the subscript $i$ stands for $\pi$, $K$ and $\eta$, $z^r$ are linear
combinations of
renormalized coefficients of ${\cal L}^{(3)}$ defined at a scale $\mu$,
$f_i (m_i)$ are calculable $\mu$-independent functions of $m_i$ and $\gamma_i$
and $\alpha_i$ are known constants. If there are enough experimental
data, one may first fix the scale $\mu$ and then determine the unknown
constants
$z_i$. This would allow the same Lagrangian with the parameters so determined
to make predictions
for other processes, the power of effective field theories. For our
purpose, this is neither feasible nor necessary. In fact, to ${\cal O}(Q^3)$ in
the chiral counting, the quantity $\bar{g}$ of eq.(\ref{scaleindep}) is
$\mu$-independent (or renormalization-group invariant). Therefore we will just
fix $\bar{g}$ directly by experiments. Of course given $\bar{g}$, we can always
re-express the first term of eq.(\ref{scaleindep}) in terms of quantities fixed
at some given $\mu$ as would be needed for comparisons with other processes.

Finally an equally important
contribution that need to be considered at ${\cal O}(Q^3)$
is the loop effect on
the property of the $\Lambda (1405)$. Because of the open channel $\Lambda
(1405)\rightarrow \Sigma\pi$, the one-loop self-energy of the $\Lambda (1405)$
becomes complex, giving an imaginary part to the $\Lambda (1405)$ mass.
If one assumes flavor $SU(3)$ symmetry which is valid at tree level ({\it
i.e.}, to ${\cal O} (Q^2)$), then
the $\Lambda (1405)\Sigma\pi$ coupling is the same as the
$\Lambda (1405)p K$ coupling, so the width corresponding to the decay
$\Lambda (1405)\rightarrow \Sigma\pi$ must be determined by the value
$\bar{g}_{\Lambda_R}^2 \approx 0.15$ obtained above.
The corresponding width comes out to be
\beq
\Gamma_{\Lambda_R}\approx 50\ \  {\rm MeV}
\eeq
which is in agreement with the empirical width of $\Lambda (1405)$.
This together with perturbative unitarity suggests that
the amplitude (\ref{TLambda}) with $m_{\Lambda_R}$ replaced by a complex mass
could be used to take into account the ${\cal O}(Q^3)$ effect.
However if we take the complex mass for $m_{\Lambda_R}$, the real part
of eq.(\ref{TLambda}) becomes smaller. Hence, in order to compensate for
the reduced amplitude, we found that it was necessary to take a larger
effective coupling, $\bar g_{\Lambda_R}^2\approx 0.25$ at ${\cal O}(Q^3)$.
This difference may perhaps be justified by the fact that
$SU(3)$ breaking which enters at one-loop order could induce
the coupling constants to differ by
as much as 30\% \cite{butler}.

The results obtained by constraining
$\bar d_s$, $\bar d_v$, $\bar g_s$ and $\bar g_v$ to the empirical data
eq. (\ref{exp}) are
\beq
\bar d_s       \approx 0.201 \; \fm, &&
\bar d_v       \approx 0.013 \; \fm, \nonumber\\
\bar g_s M_K   \approx 0.008 \; \fm, &&
\bar g_v M_K   \approx 0.002 \; \fm.
\eeq
If one takes eq.(\ref{1/mtheory}) for the $1/m_B$ corrections assuming that
higher-order $1/m_B$ corrections do not modify the fit importantly, we can
extract the parameter $\sigma_{KN}\equiv -(1/2)(\hat{m}+m_s)(a_1+2a_2+4a_3)$
needed to fit the scattering lengths
(more precisely, $\bar{d}_s\approx 0.201 \ {\rm fm}$)
\beq
\sigma_{KN}\approx 2.83 m_\pi.
\eeq
This is not {\it by itself} the $KN$ sigma term since loops renormalize it
but this enhanced value may play a role in kaon condensation phenomena.
Unfortunately complications due to the increased number of terms and off-shell
ambiguities
that are introduced at loop orders do not permit as simple an
analysis as the one made above at tree level. Further work is required
to pin down, for instance, the roles of the $1/m_B$ corrections and the $KN$
sigma term in supplying information on the strangeness content of the proton.
These complications do not, however,  obcure the  main thrust of our paper,
which is that
the attraction found at ${\cal O}(Q^2)$ in the
$KN$ interaction remains unaffected by the loop graphs and that $\langle
P|\bar{s}s|P\rangle\approx 0$ is consistent with the s-wave scattering
data.

The scattering amplitudes in each chiral order are
 given in Table 1.
One sees that while the order $Q$ and order $Q^2$ terms are comparable,
the contribution of order $Q^3$ is fairly suppressed compared with them.
As a whole, the subleading chiral corrections are verified to be
consistent with the ``naturalness" condition as required
of predictive effective field theories.
Using other sets of values of $f$, $D$ and $F$ does not change
$L_s$ and $L_v$ significantly, hence leaving unaffected  our main conclusion.

\begin{table}
$$
\begin{array}{|r||r|r|r|r|}
 \hline
{\bar g^2_{\Lambda_R}=0.25} &{\cal  O}(Q) & {\cal O}(Q^2) & {\cal O}(Q^3) &
\Lambda (1405) \\
 \hline
 \hline
a^{K^+p} (fm) & -0.588 & 0.316 & -0.114 & 0.076 \\
 \hline
a^{K^-p} (fm) &0.588 & 0.316 & -0.143 & -1.431 \\
 \hline
a^{K^+n} (fm) & -0.294 & 0.277 & -0.183 & 0.000\\
 \hline
a^{K^-n} (fm) & 0.294 & 0.277 & -0.201 & 0.000\\
 \hline
\end{array}
$$
\caption{ Scattering lengths from three leading order contributions.}
\centerline{ Also shown is the contribution from $\Lambda (1405)$.}
\end{table}

We now turn to off-shell s-wave $K^-$ forward scattering off static nucleons.
The kinematics involved are $t=0$, $q^2=q'^2=\omega^2$, $s=(m_N+\omega)^2$ with
an arbitrary (off-shell) $\omega$.
In terms of the low-energy parameters fixed by the on-shell constraints,
the off-shell $K^-N$ scattering amplitude\footnote{The off-shell amplitude
calculated here does not satisfy Adler's soft-meson conditions that follow from
the usual PCAC assumption that the pseudoscalar meson field interpolate as the
divergence of the axial current. This is because the meson fields of the chiral
Lagrangian do not interpolate in the same way in the presence of
explicit chiral symmetry breaking. However one can always redefine the meson
fields consistently with chiral symmetry
without changing the S-matrix so as to recover Adler's conditions.
(Specifically this can be assured by imposing external gauge invariance in
(pseudo-) scalar channel.) To the
same chiral order, therefore,
the two different (and all other equivalent)
off-shell amplitudes have not only the same on-shell limit but perhaps also the
same effective action at {\it the extremum point}. We do not have a rigorous
proof of this statement but we believe it to be reasonable to assume
that all physical quantities (including equation of state) computed from such
an effective action would not depend upon how the meson field extrapolates
off-shell, which is of course at variance with the arguments made by the
authors in \cite{YNMK,lutz}. This point was stressed to us by Aneesh Manohar.}
as a function of $\omega$ comes out to be
\beq
 a^{K^-p} &=&\left.  \frac{1}{4\pi (1+\omega/m_N)} \right\{
   T_v^{K^-p}(\omega=M_K)
  - \frac{\omega^2}{f^2}
     \left(\frac{\bar g_{\Lambda_R}^2}{\omega+m_B-m_{\Lambda_R}} \right)
\nonumber\\ && \;\;\;\;\;\;\;\;\;\;
 +\frac{1}{f^2} (\omega-M_K) +\frac{1}{f^2} (\omega^2-M_K^2)
\left(\bar d_s-\frac{\sigma_{KN}}{M_K^2} +\bar d_v +\frac{(\hat m+m_s)
a_1}{2M_K^2}\right)
\nonumber\\ && \;\;\;\;\;\;\;\;\;\; \left.
   +\frac{1}{f^2} ( L^+_p (\omega) -L^+_p(M_K))
 -\frac{1}{f^2} ( L^-_p (\omega) -L^-_p (M_K))  \right\},\\
 a^{K^-n} &=& \left. \frac{1}{4\pi (1+\omega/m_N)} \right\{
 T_v^{K^-n}(\omega=M_K)
\nonumber\\ && \;\;\;\;\;\;\;\;\;\;
 \frac{1}{2f^2} (\omega-M_K)
 +\frac{1}{f^2}({\omega}^2 -M_K^2)
\left(\bar d_s -\frac{\sigma_{KN}}{M_K^2} -\bar d_v
  -\frac{(\hat m+m_s)a_1}{2M_K^2}\right)
\nonumber\\ && \;\;\;\;\;\;\;\;\;\; \left.
  + \frac{1}{f^2}(L^+_n (\omega) -L^+_n(M_K))
 -\frac{1}{f^2} (L^-_n(\omega)-L^-_n(M_K))   \right\}.
\eeq
Here\footnote{The functions $L_{p,n}^-(\omega)$ contain four parameters
$\alpha_{p,n}$ and $\beta_{p,n}$. Owing to the constraints at $\omega=M_K$,
$L^-_{p,n} (M_K)$,
they reduce to two. These two cannot be fixed by on-shell data.
However since the off-shell amplitudes are
rather insensitive to the precise values of these constants, we will somewhat
arbitrarily set $\alpha_{p,n} \approx \beta_{p,n}$ in calculating Figure 2.}
\beq
L^+_p(\omega)
  &=&\frac{\omega^2}{64\pi f^2} \Big\{
  \Big[2(D-F)^2+\frac 13(D+3F)^2\Big] M_K
+ \frac{3}{2} (D+F)^2 M_\pi
\nonumber\\ &&  \;\;\;\;\; \;\;\;\;\;
+\frac 12 (D-3F)^2 M_\eta
-\frac {1}{3}(D+F)(D-3F)( M_\pi+M_\eta)
\nonumber\\
&& \;\;\;\;\; \;\;\;\;\;
-\frac {1}{6}(D+F)(D-3F) (M_\pi^2+M_\eta^2)\int^1_0 dx \frac{1}{
 \sqrt{(1-x)M_\pi^2+ xM_\eta^2}}  \Big\} \nonumber\\
&& +\frac{\omega^2}{8 f^2}
  \left(4\Sigma_K^{(+)}(-\omega)+  5\Sigma_K^{(+)}(\omega)
   +2\Sigma_\pi^{(+)}(\omega)
+3\Sigma_\eta^{(+)}(\omega)\right),
\nonumber\\
L^-_p(\omega) &=& \alpha_p M_K^2 \omega +\beta_p \omega^3
 +\frac{1}{4f^2} \omega^2 \left\{-\frac 12\Sigma_K^{(-)}(\omega)
 -\Sigma_\pi^{(-)}(\omega)-\frac 32\Sigma_\eta^{(-)}(\omega)\right\},
\nonumber\\
L^-_p(M_K) &=& (\bar g_s+\bar g_v) M_K^3,\nonumber\\
L^+_n(\omega)
  &=&\frac{1}{64\pi f^2}\omega^2 \Big\{
  \Big[\frac 52 (D-F)^2+\frac 16(D+3F)^2\Big] M_K
+  \frac 32 (D+F)^2 M_\pi
 \nonumber\\
&& \;\;\;\;\; \;\;\;\;\;
+\frac 12 (D-3F)^2 M_\eta
 +\frac 13 (D+F)(D-3F)(M_\pi+M_\eta)
\nonumber\\ && \;\;\;\;\; \;\;\;\;\;
 +\frac 16 (D+F)(D-3F)(M_\pi^2+M_\eta^2) \int^1_0 dx \frac{1}{
 \sqrt{(1-x)M_\pi^2+x M_\eta^2}} \Big\} \nonumber\\
&&+ \frac{\omega^2}{8 f^2} \cdot
  \left(2\Sigma_K^{(+)}(-\omega)+
   \Sigma_K^{(+)}(\omega) +\frac 52 \Sigma_\pi^{(+)}(\omega)
 +\frac 32 \Sigma_\eta^{(+)}(\omega) \right),
\nonumber\\
L^-_n(\omega) &=& \alpha_n M_K^2 \omega +\beta_n \omega^3
 +\frac{1}{4f^2} \omega^2 \left\{\frac 12\Sigma_K^{(-)}(\omega)
 -\frac 54\Sigma_\pi^{(-)}(\omega)-\frac 34\Sigma_\eta^{(-)}(\omega)\right\},
\nonumber\\
L_n^-(M_K) &=& (\bar g_s-\bar g_v) M_K^3,
\eeq
where
\beq
\Sigma_i^{(+)} (\omega)
  &=& -\frac{1}{4\pi} \sqrt{M_i^2-\omega^2} \times \theta(M_i-|\omega|)
  +\frac{i}{2\pi}\sqrt{\omega^2-M_i^2}\times \theta(\omega-M_i),
\nonumber\\
\Sigma_i^{(-)} (\omega)&=&
  -\frac{1}{4\pi^2} \sqrt{\omega^2-M_i^2}\ln\left|\frac{
  \omega+\sqrt{\omega^2-M_i^2}}{\omega-\sqrt{\omega^2-M_i^2}}\right|
\times   \theta (|\omega|-M_i)
\nonumber\\
&&  -\frac{1}{2\pi^2}\sqrt{M_i^2-\omega^2}\sin^{-1}
\frac{\omega}{M_i}\times \theta(M_i-|\omega|).
\eeq
The results for $K^- p$ and $K^-n$ scattering are summarized in Figure 2
for the range of $\sqrt{s}$ from $1.3$ GeV to $1.5$ GeV with
 ${\bar g}_{\Lambda_R}^2=0.25$ and $\Gamma_{\Lambda_R}=50$ MeV.
Those for $K^- n$ scattering are independent of the $\Lambda (1405)$ and
vary smoothly over the range involved.
 Our predicted $K^-p$ amplitude is found to be in fairly good agreement with
the
 recent fit by  Steiner\cite{Steiner}.
The striking feature of the real part of the $K^-p$ amplitude,
repulsive above and attractive
below $m_\Lambda(1405 MeV)$ as observed here, and the $\omega$-independent
attraction of the $K^- n$ amplitude are quite possibly relevant to kaonic atoms
\cite{gal} and to kaon condensation in ``nuclear star" matter. In comparison
with Steiner's results, the imaginary part of the $K^- p$
amplitude predicted here is a bit too big. This may be due to
our approximation of putting the experimental decay width of
$\Lambda (1405)$ for the imaginary part  of its mass.

The calculation of the effective action at loop orders
needed for kaonic atom and kaon condensation phenomena
will be reported elsewhere.
\vspace{0.3in}

\centerline{\large \bf Acknowledgments}

We are grateful for valuable discussions with Gerry Brown, Aneesh Manohar,
Maciej Nowak and Tae-Sun Park.
We would like to also thank Andreas Steiner for providing us with his
results on $K^- N$ amplitudes. Part of this work was done while three of us
(CHL, HJ and MR) were participating in the Fall 93 Workshop on ``Chiral
Symmetry
in Hadrons and Nuclei" at the European Centre for Theoretical Studies in
Nuclear
Physics and Related Areas (ECT$^*$), Trento, Italy. We acknowledge the
hospitality and exciting working conditions provided by the ECT$^*$ staff.
The work of CHL, HJ and DPM was supported in part by the Korea Science and
Engineering Foundation through the Center for Theoretical Physics of Seoul
National University and that of MR by the US Department of Energy under Grant
No. DE-FG02-88ER40388.

\newpage
\centerline{\bf Figure Captions}
\begin{itemize}
\item {\bf Figure 1:} One-loop Feynman diagrams contributing to $K^\pm N$
scattering: The solid line represents baryons (nucleon for the external and
octet and decuplet baryons for the internal line) and the broken line
pseudo-Goldstone bosons ($K^\pm$ for the external and $K$, $\pi$ and $\eta$ for
the internal line). There are in total thirteen diagrams at one loop, but for
s-wave $KN$ scattering, for reasons described in the text, we are left with
only six topologically distinct one-loop diagrams.
\item {\bf Figure 2:} $K^-N$ amplitudes as function of $\sqrt{s}$: These
figures correspond to eqs.(23) and (24) with $\bar{g}_{\Lambda_R}^2=0.25$,
$\Gamma_{\Lambda_R}=50$ MeV and $\alpha_{p,n}=\beta_{p,n}$, fixed in the way
described in the text. The first kink corresponds to the $KN$ threshold and the
second around 1.5 GeV to $\sqrt s=m_N+ M_\eta$ for $M_\eta\approx 547$ MeV.
\end{itemize}
\end{document}